\documentclass[12pt]{article}
\usepackage{color}
\usepackage{latexsym}
\usepackage{amsmath}
\usepackage{amsfonts}
\usepackage{color}
\usepackage{graphicx}
\usepackage{amssymb}
\usepackage{indentfirst} 
\numberwithin{equation}{section}
\newcommand{\be}{\begin{equation}}
\newcommand{\ee}{\end{equation}}

\newcommand{\mN}{{\mathbb N}}

\newcommand{\tr}{\textrm{tr}}

\newcommand{\bx}{{\bf{x}}}
\newcommand{\bp}{{\bf{p}}}

\newcommand\braket[2]{\langle{#1}|{#2}\rangle}
\title{Dynamics of  entropy and information  of  time-dependent quantum systems:    exact results }
\author{
\\ 
\vspace*{0.3cm}
K. Andrzejewski\footnote{e-mail: krzysztof.andrzejewski@uni.lodz.pl}
 \vspace*{0.3cm}
\\
\small  Faculty of Physics and Applied Informatics  \\ \small
 University of  Lodz\\ \small Pomorska 149/153,
90-236, Lodz, Poland\\
}
\date{}
\begin{document}
\maketitle 
\begin{abstract} 
Dynamical aspects of  information-theoretic  and entropic measures of  quantum systems are studied. First,   we show that    for  the   time-dependent harmonic oscillator, as well as  for the charged particle in certain  time-varying electromagnetic fields, the increase of the entropy and dynamics of the Fisher information   can be directly  described and related. To illustrate  these results   we have  considered  several examples  for which all the  relations take the elementary  form.  Moreover,  we  show that the integrals of   (geodesic) motion  associated  with  some conformal Killing vectors   lead to the Ermakov-Lewis invariants   for   the considered electromagnetic  fields. Next, we  explicitly work out   the  dynamics of the entanglement entropy of the   oscillators coupled by a continuous  time-dependent parameter   as well as  we analyse some aspects of  quantum-classical transition  (in particular decoherence).   Finally,    we study  in some detail the   behavior of quantum quenches (in the presence of the critical points) for the case of   mutually non-interacting non-relativistic fermions in a harmonic trap. 
\end{abstract} 
\newpage 
\section{Introduction}
\label{s0}
The study of the information-theoretic and entropic  aspects  of quantum systems has attracted considerable interest in the recent  years. Apart from their  basic applications in  quantum information processing  (or even technology) they are also  relevant   for non-equilibrium phenomena and other branches of physics. In order to describe  the intrinsic ``uncertainty"  of  the quantum states, various  information-theoretic   measures  have been proposed;  among others the Shannon \cite{b1} (in general  R\'enyi \cite{b2}) entropy and the Fisher information \cite{b3} are the most popular.  The Shannon entropy  applied first  to the study of  fundamental limits on signal
processing operations, on the quantum level was  related to the uncertainty of the particle position (delocalization); it   leads also  to uncertainty entropic relations (alternatives for the classical  Heisenberg uncertainty relation) \cite{b4}.   In contrast to this, the  Fisher information,  which arose  from the   statistical estimation theory,  provides a more   local description of  uncertainty (it contains the gradient of the density, thus it is  more  sensitive to the local oscillations); despite  of these differences it  can also lead   to some uncertainty relations \cite{b5} as well as is related to  other measures  (for example,  by the Cram\'er-Rao inequalities \cite{b6,b7}). Finally, there are  some composite measures like the   Fisher-Shannon products \cite{b7a}-\cite{b7c}. 
\par     On the other hand, the notion of the      von Neumann entropy as well as its generalization the R\'enyi entanglement entropy (see, e.g.,  the review  \cite{b8})   plays the prominent role in the characterization  of  the quantum entanglement which, in turn,  is crucial for  quantum information processing  and makes quantum computers so  tempting.  Finally, the entanglement entropy appears in other contexts such as  non-equilibrium processes, many-body physics and cosmology.
\par
In view of the  above the dynamical properties of the mentioned  measures seem very  interesting and have been studied extensively. Various time-dependent systems  were analysed  and the evolution of  information measures discussed.  Let us mention here a few of them in the context of the results presented in this paper.  
\par 
 One of  the basic examples of the time-dependent system is the harmonic oscillator with time-dependent frequency. Quantum dynamics of such a system have been  studied since the classical works \cite{b9a,b9b}.  It turns out that  the evolution of the quantum states can be reduced to the solutions of the classical Ermakov-Milne-Pinney  (EMP)  equation \cite{b10a,b10b,b10c};  in turn  such a relation   can  simplify considerations, see the preliminary section \ref{s1a}.  The Shannon entropy  for such a system  has been  considered in Ref. \cite{b11}.  
Some other information measures  (including the joint entropy and the Fisher information)  for special  cases of frequencies   and states (mainly related to the ground state) were analysed   in Refs. \cite{b12a}-\cite{b12e}. In the present   work, see  Sec. \ref{s1c},  we  note that despite   the fact that  the explicit form of the Shannon,  joint  and R\'enyi entropies    is not directly  accessible, their increase can be easily described for a whole basis of states. Moreover, we will show that the evolution of the  position and momentum Fisher informations can be directly   computed also  for various excited states what enables  their further examination in the context of the uncertainty relations.   	 To make these results as visible as possible first, in Sec. \ref{s1b},  we give some examples of frequencies for which the EMP equation is  solvable and, what is more, the solutions can be  expressed in terms of elementary functions.   
\par
Another  natural example of  the  time-dependent  system  is  the  charged particle in  time-varying  electromagnetic fields.   It turns out that for some cases  of fields we can find the basic solutions of the Schr\"odinger  equation what enables further analysis of  information-theoretic aspects. Such a situation has been discussed  in Ref. \cite{b13}  for  the particle  which  is initially  in the  ground state.  
 In Sec. \ref{s2a}   we show that these considerations can be extended  for more states if we change the basis of the solutions of the Schr\"odinger equation;    we  construct also the time-independent Fisher-Shannon
complexities for the new basis.
Moreover, the  above mentioned  special  frequencies  can be used to  construct electromagnetic pulses (including the Dirac delta behavior)  for which information-theoretic considerations take a simple form.  We  also note, see Sec. \ref{s2b}, that  such a  solvability can be related to the  conformal symmetry. To this end, first,  by means   of the so-called Eisenhart-Duval lift \cite{b13a}-\cite{b13e} we show that  some conformal Killing fields  lead to Ermakov-Lewis   invariants for  electromagnetic fields.   
\par
Next, we   analyse some examples of  the   dynamics of   the entanglement entropy of states of continuous variables.  More precisely, we  consider  the system of  two coupled  oscillators. Then the dynamics of a subsystem  consisting of the one oscillator (in the bipartite decomposition of the total system)  is described by the reduced density operator.   When the coupling parameter (optionally, together  with  frequencies)  is  time-dependent then we  observe time-dependence of the  entanglement entropy.  For the von Neumann (R\'enyi) entropy and instantaneous (infinitely-fast)  quenches  this problem has been studied in Ref. \cite{b14}.  
In addition, in Sec. \ref{s3a},   we construct   exactly solvable models  of   continuous time-dependent coupling  parameter   which  enable  us to perform  an  elementary  analysis of   the dynamics of the entanglement entropy,  complementing in this way  the discussion presented in Refs. \cite{b14};  in particular, we give an example when the final value of  the entanglement entropy stabilizes independently of
the history of the evolution. 
\par
Moreover, in Sec. \ref{s3b} we analyse some aspects of the transition  from quantum to classical  world  which is important for many branches of physics:   starting from   quantum measurements through   condensed matter physics,    open system and ending with quantum gravity and cosmology  (let us only mention   here a few references   \cite{br1a}-\cite{br1d}). One of the main aspects of this transition is the loss of coherence.  It has attracted  increasing   interest  in the last years   due to the  great  importance of   the (de)coherence phenomena  for quantum computation and quantum  telecommunication (see  \cite{br1b,br2a,br2b,br2c} and references therein); the   interaction of the system  with the  surrounding  can result in  loss of the quantum properties (in particular the quantum entanglement).  This is  especially important for the quantum memory which should be the faithful storage of  quantum  information (see, e.g., \cite{br3a}). This problem has been  recognized from the very  beginning  \cite{br3b} and various error correction methods proposed. However,  to ensure the quantum error corrections, for  large-scale quantum computations, decoherence effects  on quantum gates should be reduced   (especially if we take into account the fact that more and more  operations  can accumulate decoherence).  In view of this the understanding of the mechanisms of decoherence  is the  pivotal  problem  which  has been studied in various ways and models. Here, we analyse these issues,  by means of the  models described above, for  two  measures of the classicality: quantum decoherence (related to the damping of the off-diagonal elements of the density matrix) and  the so-called classical correlation (basing on  the form of the  Wigner function), see  \cite{br1a,br1d}. To this end we use, in particular,  the model of  coupled oscillators to simulate  the time-dependent interaction of the  systems with the environment. 
\par
The notion of the entanglement entropy  appears  also in relation to a pure quantum state confined to some region  \cite{b15a,b15b} of space or boundary between two parts of a quantum many-body system, e.g.  \cite{b16a,b16b}.  Quite recently,  such  a situation   has been  discussed in Ref.  \cite{b17}  for the  entanglement entropy of a  given subregion of  the   system of many   non-relativistic fermions in external  time-dependent harmonic  traps.  This  model is interesting due the quantum field theory description of  non-equilibrium and  critical  phenomena.   In particular, it has been shown that for large number of fermions the entanglement entropy of a subregion and basic expectation values are also determined by the solution of the EMP equation.  Thus, in Sec. \ref{s4} we use  a special form of the frequencies to simplify discussion of (a)diabatic  phenomena which appears in the presence  of a  critical point.
\par
Finlay,  in  Sec.  \ref{s5} we summarize all the results    obtained     as well as  we outline further directions of investigations.  
\section{Dynamics of  entropy and information for \\ time-dependent oscillator}
\label{s1}
\subsection {Preliminaries }
\label{s1a}
Let us  start with  the classical  harmonic  oscillator  with the   time-dependent frequency  $\omega(t)$. It is described  by the  Hamiltonian
\be
\label{e1}
H(t)=\frac{p^2}{2}+\frac{1}{2}\omega^2(t)x^2,
\ee
for which the  equation of motion reads
\be
\label{e2}
\overset{..}{x}(t)=-\omega^2(t)x(t).
\ee
 The time-dependent harmonic oscillator (TDHO)  appears in many physical models and has been studied in various contexts.  It turns out  that  such a  system   is equivalent  to   the so-called  Ermakov-Milne-Pinney (EMP) equation \cite{b10a,b10b,b10c} 
 \be
 \label{e3}
\overset{..}{b}(t)+\omega^2(t)b(t)=\frac{c^2}{b^3(t)},
\ee 
where $c$ is a constant (we assume  $c\neq 0 $ to ensure the non-vanishing of the function $b(t)$). 
In fact,  let  $x_1(t)$ and $x_2(t)$ be  two linearly independent solutions to eq. \eqref{e2}  and  $W$ the Wronskian of $x_1(t)$ and $x_2(t)$   then 
\be
\label{e4}
b(t)=\sqrt{x_1^2(t)+\frac{c^2}{W^2} x_2^2(t)},
\ee
is a solution of eq.  \eqref{e3}. Conversely, if $b(t)$ is a solution to eq. \eqref{e3}   then the real and  imaginary parts  of the function 
\be
\label{e5}
B(t)=\frac{i}{\sqrt{2c}}b(t)e^{ic\tau(t)},
\ee
where 
\be
\label{e6}
\tau(t)=\int \frac {dt}{b^2(t)},
\ee
form a fundamental set of  solutions of eq. \eqref{e2}; moreover, the following identity holds 
\be
\label{6a}
\dot B(t)\bar B(t)-\dot{\bar B}(t)B(t)=i.
\ee
Although equation \eqref{e3}  seems more complicated  than the initial one (it is a non-linear one) the function $b(t)$ has a nice interpretation. Namely,    the transformation
 \be
 \label{e7}
 y=\frac{ x} {b(t)} \quad  \tau=\tau(t),
 \ee
 to the new coordinate $y$ and time $\tau$  maps eq. \eqref{e2}   into the harmonic oscillator equation 
\be
\label{e8}
{y}''(\tau)=-c^2y(\tau),
\ee
(prime denotes the derivative w.r.t to $\tau$) for which the solutions are well-known;  in consequence, the equivalence of  both equations \eqref{e2} and \eqref{e3}  is now more clear.  
\vspace{0.5cm}
\par
Since equation  \eqref{e2}  is a  linear one,  one can expect that a  similar situation appears also at  the quantum level.  In fact, it turns out that the  dynamics  of the quantum TDHO  can be reduced, remarkably  by means of  the function $b(t)$, to the ordinary  quantum harmonic oscillator.  This fact can be  observed is several ways.  It seems  that the most direct approach  is based on  the observation (see \cite{b18})  that the transformation \eqref{e7} can be lifted  to a unitary  transformation
\be
\label{e9}
\psi(x,t)=\frac{1}{\sqrt{ b(t)}}\phi({x}/{b(t)},\tau(t))e^{\frac{i\dot b(t)x^2}{2b(t)}},
\ee
which maps  the solution $\phi(y,\tau)$ of the   Schr\"odinger equation with  the   Hamiltonian operator 
\be
\label{e10}
\hat H_y=\frac {\hat p_y^2}{2}+\frac{c^2\hat y^2}{2},
\ee
to the solution $\psi(x,t)$ of the Schr\"odinger  equation with the Hamiltonian \eqref{e1}. 
It is worth to notice  that we can go  even further  and relate $\phi$  to the solution $\tilde \phi $ of the  free particle   by means of the   the so-called Niederer transformation, see \cite{b19a} (for more recent details of this issue see  \cite{b19b})
\be
\label{e10b}
\phi( {\bf y},t)= \frac{e^{-\frac{ic}{2}\tan(ct) {\bf y}^2}}{\cos(ct)}\tilde \phi\left (\frac{ {\bf y}}{\cos(ct )},\frac  {\tan(ct)}{c}\right);
\ee
however, the price we pay is that  the transformation \eqref{e10b} is  a local one. 
\par 
In view of the   transformation  \eqref{e9}  we immediately obtain   that the general solution for the Schr\"odinger equation of the TDHO is  a superposition of the following wave functions 
\be
\label{e11}
\psi_n(x,t)=\frac{1}{\sqrt{2^nn! b(t)}}\sqrt[4]{\frac{c}{\pi}}e^{-ic(n+1/2)\tau(t)}H_n\left(\frac{\sqrt{c} x}{ b(t)}\right)e^{-\frac{cx^2}{2b^2(t)}+\frac{i\dot b(t)x^2}{2b(t)}},
\ee
where $H_n$ for $n=0,1\ldots$ denote the Hermite polynomials; or equivalently,  in terms of the function $B(t)$  they are given by 
\be
\label{e12}
\psi_n(x,t)=\frac{1}{\sqrt{2^nn! |B(t)|}}\frac{1}{\sqrt[4]{2\pi}}\left(-\frac{\bar B(t)}{B(t)}\right)^{(n+1/2)/2}H_n\left(\frac{ x}{\sqrt 2 |B(t)|}\right)e^{\frac{i\dot B(t)}{2B(t)}x^2}.
\ee
 \par 
Another  way  to see the discussed  relation  is   based on the conserved quantities.  In this approach we start with  the   Hamiltonian operator  \eqref{e10} of the harmonic oscillator    which  is $\tau$ independent.  Then, by means of the transformation \eqref{e7},  one obtains the Lewis-Riesenfeld (LR)   operator 
\be
\label{e13}
\hat I(t)=\frac{1}{2}\left(\frac{c^2\hat x^2}{b^2(t)}+(\hat pb(t)-\hat x\dot b(t))^2\right),
\ee
which satisfies the quantum Liouville-von Neumann equation: $i\partial_t\hat I+[\hat I,\hat H]=0$, so it is a constant of motion (i.e., its  mean values do not depend on time for any state obeying the Schr\"odinger equation).
The same concerns the $\tau$ dependent  annihilation   (and creation) operator  $\hat a(\tau) =e^{ic\tau}\hat a$ which   after transformation \eqref{e7} takes the form
\be
\label{e14}
\hat a(t)=\frac{e^{ic\tau(t)}}{\sqrt{2}}\left(\frac {\sqrt{c}\hat x}{ b(t)}+\frac{i}{\sqrt{c}}(\hat pb(t)-\hat x\dot b(t))\right)=B(t)\hat p-\dot B(t)\hat x,
\ee
(and similarly  for $\hat a^\dag$(t)).  As a consistency check let us note that     $\hat I(t)=c(\hat a^\dag(t)\hat a(t)+1/2)$ and $[\hat a(t),\hat a^\dag(t)]=1$. Now, following the Lewis and Riesenfeld    observation  \cite{b9b} the eigenfunctions of the operator $\hat I$  are, up to a  time-depend phase,   solutions  to the Schr\"odinger equation for the TDHO. In turn,  the latter ones  can be   found  by means of $\hat a(t)$ and  $\hat a^\dag(t)$ operators while   the phase correction, for example,    by the direct substitution  to the Schr\"odinger equation. In consequence,    we arrive  at    the  desired states \eqref{e11}.  
\par
 In the third  approach,  we  construct the Fock space corresponding  to the  operators $\hat a(t)$ and $\hat a^\dag(t)$ in the position representation. Namely, the states $\tilde  \psi_n(x,t)=\braket{x}{n,t}$  are obtained by the well-known  formula  $\tilde  \psi_n(x,t)=\frac{1}{\sqrt{n!}}(\hat a^\dag(t))^n\tilde \psi_0(x,t)$  where $\hat a(t)\tilde \psi_0(x,t)=0$ and $\hat a (t)$ is given by \eqref{e14}. Again, after some computations (see e.g.  \cite{b20})  one obtains that  the states  $\tilde \psi_n(x,t)$ coincide  with  \eqref{e11}  (equivalently   \eqref{e12}) up to a time-dependent phase which   can be found in  the same way as above  (of course this phase  correction can be eliminated from the very beginning, since the state $\tilde\psi_0(x,t)$  is defined modulo a time-depend phase).
 \vspace{0.5cm}
 \par
 Sometimes, in  physical considerations  we  want to analyse the  dynamics of the   state which at  initial time $t=t_0$ is an  eigenstate  of the instantaneous Hamiltonian $\hat H(t_0)$.   For the quantum TDHO, by the inspection of eq. \eqref{e11},  we see that this holds when we    put 
\be
\label{e15}
 c=\omega(t_0)\equiv\omega_0,\quad  b(t_0)=1, \quad \dot b(t_0)=0.
\ee
 Equivalently, in terms of $x_1(t)$ and $x_2(t)$ (see  eq. \eqref{e4})     $x_1(t_0)=1, \dot{x}_1(t_0)=0,\quad   x_2(t_0)=0$   or,  in terms of $B(t)$ (see eq. \eqref{e5}) $B(t_0)=\frac{i}{\sqrt{2\omega_0}}$ and $\dot B(t_0)=-\sqrt{\frac{\omega_0}{2}} $ (the last ones coincide with  the identity  \eqref{6a}).
 \par 
 In  view of the above discussion  the dynamics of  states  of the quantum TDHO (and consequently various   physical systems which can be reduced to it) is related to   the function $b(t)$  satisfying eq.  \eqref{e3}. 
 In consequence,       many  interesting quantities can be   expressed in terms of this function.  A few popular ones, for the basis $\psi_n(x,t)$,  take  the form 
 \be
 \label{e16}
 \langle  x^2\rangle _n(t)=\frac{b^2(t)}{2\omega_0}(2n+1), \quad \langle  p^2\rangle _n(t)=\left(\frac{\omega_0}{b^2(t)}+\frac{\dot b^2(t)}{\omega_0}\right)\frac{(2n+  1)}{2};
  \ee
 \be
 \label{e17}
 \langle  H\rangle _n(t)= \frac 14 \left (\frac{\omega^2 (t)b^2(t)}{\omega_0}+\frac{ \omega_0}{ b^2(t)}+\frac{\dot b^2(t)}{\omega_0}\right)(2n+ 1),  \quad \Delta_n x(t) \Delta_n p(t)=(n+\frac 12)\sqrt{1+\frac{b^2(t)\dot b^2(t)}{\omega_0^2}}
 \ee
 \subsection{Explicit examples}
 \label{s1b}
As we have noted in the  previous section in the analysis of the dynamics of the quantum TDHO  the solutions to  equation \eqref{e3} are   crucial.  Of course, the are some special frequencies  when the explicit  form of the function $b(t)$ is known. For discontinuous $\omega(t) $  the  most popular  is the so-called abrupt  profile when the   frequency  is instantly changed from  one value to another one. For the smooth   $\omega(t)$ the situation is more complicated and  the profiles basing   on the hyperbolic tangent  function are frequently used; then, however,  the function $b(t)$ is given by some special functions, which in turn are difficult in   a further analysis.   Here,  we analyse  some special choices of the frequency  for which the function $b(t)$ is an elementary one; in consequence, the analysis of physically interesting quantities can be simplified.  
\par
  To this end  let us consider the following  family of the frequencies  
\be
\label{e18}
\omega_{I}^2(t)=\frac{2}{\epsilon^2\cosh^2(t/\epsilon)}+a^2;
\ee
then   $\omega_{I}(t)$is a  bell shaped function with the maximum at $t=0$ and   the same initial and final value $a^2$ (in general non-zero). 
\par 
For the case $a>0$   we consider  two useful  initial conditions. First,  we  take  the initial conditions \eqref{e15} with  $t_0=0$ (i.e. at the maximum).  Then  we have 
\be
\label{e20}
b^2(t)=\left(1+\frac{\tanh^2(t/\epsilon )}{a^2\epsilon ^2}\right)\left(1-\frac{\sin^2\left(at+\tan^{-1}(\frac{\tanh(t/\epsilon)}{a\epsilon})\right)}{(1+a^2\epsilon^2)^2}\right).
\ee
Moreover, the function $\tau(t)$ (see eq. \eqref{e6})  can be also explicitly computed
 \be
 \label{e21} 
 \tau(t)=\frac{\epsilon}{\sqrt{2+a^2\epsilon^2}}\tan^{-1}\left(\frac{a\epsilon\sqrt{2+a^2\epsilon^2}}{1+a^2\epsilon^2}\tan\left(at+\tan^{-1}\left(\frac{\tanh(t/\epsilon)}{a\epsilon}\right)\right)\right).
 \ee
  For $a=0$   the functions $b(t)$ and $\tau(t)$ can be found  directly or by taking a careful limit of eqs. \eqref{e20} and \eqref{e21}; for example, one gets
\be
\label{e19}
b^2(t)=\left(1-\frac{t}{\epsilon}\tanh(t/\epsilon)\right)^2+2\tanh^2(t/\epsilon).
\ee
\par In order to obtain  quenched models   we  define   $\tilde\omega_{I}(t)$ as follows
 \be
 \label{e22}
\tilde \omega_{I}^2(t)= \left\{
\begin{array}{cc}
\omega_0^2\equiv  a^2+2/\epsilon^2 & \quad \textrm{for} \quad  t\leq 0,\\ 
\omega_{I}^2(t)& \quad \textrm{for} \quad  0< t.
\end{array}
\right.
\ee  
Then   the initial value $a^2+2/\epsilon^2$  is quenched to  $a^2$ and  the corresponding  function $\tilde b(t) $  reads
 \be
\label{e31}
\tilde b(t)= \left\{
\begin{array}{cc}
1 & \quad \textrm{for} \quad  t\leq 0,\\ 
b(t)& \quad \textrm{for} \quad 0<t;
\end{array}
\right.
\ee
  where $b(t)$ is given by \eqref{e20}  or by \eqref{e19} for $a>0$ or $a=0$, respectively.  
 \par
 The second interesting  initial condition is given by \eqref{e15} with $t_0=-\infty$. Then the function $b(t)$ is given by
 \be
 \label{e23}
 b^2(t)=\frac{a^2\epsilon^2+\tanh^2(t/\epsilon)}{1+a^2\epsilon^2},
 \ee
 while 
\be
\label{e24}
 \tau(t)=t+\frac{1}{a}\tan^{-1}\left(\frac{\tanh(t/\epsilon)}{a\epsilon}\right).
 \ee
 A remarkable property of this case is that the function \eqref{e23}    satisfies  the condition \eqref{e15} also at $t=\infty$ (there is no oscillatory behavior  for  $a>0$ at plus infinity). Thus  the state $\psi_n$   at $t=\infty$  is  again an eigenstate of the instantaneous  Hamiltonian  operator $\hat H(\infty)=\frac 12(p^2+a^2x^2)$.  What is more,  this  fact is independent of the parameter $\epsilon$  which controls  the maximal value of the frequency  $\omega_{I}$ (in other words, independently of   the history of the evolution).
 \vspace{0.5cm}
 \par 
The  above family of the  frequencies will  be  useful  for illustrating  our  further considerations; however,  it    does not have a parameter which control  the slope rate  (e.g., from $\omega_{I}^2(0)$  to $\omega_{I}^2(\infty)$), $\epsilon$ controls the maximal value of the frequency. Such a possibility  is relevant  for some investigations; for example, we  cannot perform  the abrupt limit and investigate other (a)diabatic properties.  In order to improve this situation let us introduce the second family of frequencies defined as follows
 \be
 \label{e25}
\omega_{II}(t)=\frac{a}{t^2+\epsilon^2},
\ee
where $a,\epsilon>0$; such a profile exhibits also the bell   shape with the maximum  at $t=0$ but  this time the parameter $\epsilon$ controls the slope rate.  The general   solution of eq. \eqref {e2} is of the form 
\be
\label{e26}
\begin{split}
x=C_0\sqrt{t^2+\epsilon^2}\cos(\lambda \tan^{-1}(t/\epsilon)+C_1),
\end{split}
\ee
where $\lambda^2=1+\frac{a^2}{\epsilon^2}$.   Thus by virtue of eq. \eqref{e4} we can easily find  the function $b(t)$. 
\par 
Namely,   the initial conditions \eqref{e15} imply the following form of $b(t)$
\be
\label{e27}
b^2(t)=\frac{(t^2+\epsilon^2)}{\lambda^2\epsilon^2(t_0^2+\epsilon^2)}\left((\epsilon^2+a^2+t^2_0)\cos^2(\lambda \tan^{-1}(t/\epsilon)+C)+a^2\sin^2(\lambda\tan^{-1}(t/\epsilon)+\tilde C)\right ),
\ee
where $C=\tilde C-\tan^{-1}(t_0/(\lambda\epsilon))$ and  $\tilde C=\lambda \tan^{-1}(t_0/\epsilon)$.
Moreover, the function $\tau(t)$ can be also explicitly found
  \be
  \label{e28}
  \tau(t)=\frac{t_0^2+\epsilon^2}{a}\tan^{-1}\left(\frac{t_0}{a}+\frac{t_0^2+a^2}{a\sqrt{a^2+\epsilon^2}}\tan(\lambda\tan^{-1}(t/\epsilon)+\tilde  C)\right).
  \ee
In  particular,  taking  $t_0=0$  (i.e.  the point where is the maximum of the frequency) we have 
\be
\label{e28b}
b^2(t)=\frac{(t^2+\epsilon^2)}{a^2+\epsilon^2}\left(\cos^2(\lambda \tan^{-1}(t/\epsilon))+\frac{a^2}{\epsilon^2}\right).
\ee
Furthermore,  for the specific value $a=\sqrt{(k^2-1)}\epsilon$, $k\in N$  the right-hand side of eq.  \eqref{e28b} reduces to  a rational  function; for example, taking   $a=\sqrt{3}\epsilon$ ($k=2$) one obtains 
\be
\label{e29}
b^2(t)=1+\frac{t^4}{\epsilon^2(t^2+\epsilon^2)}.
\ee
 \vspace{0.5cm}
\par 
In contrast to   $\omega_I(t)$ described above  the frequency $\omega_{II}(t)$ tends to zero at infinities.  We   can  change this situation   by considering the   profile of the form 
\be
\label{e30}
\tilde \omega_{II}(t)= \left\{
\begin{array}{cc}
\omega_0\equiv \frac{a}{\epsilon^2+t_0^2} & \quad \textrm{for} \quad  t\leq t_0,\\ 
\omega_{II}(t)& \quad \textrm{for} \quad  t_0< t;
\end{array}
\right.
\ee  
 then the solution  of eq. \eqref{e3}  with the initial conditions \eqref{e15}   can be obtained in  a similar way as in eq.   \eqref{e31}.
In particular,  taking  in eq. \eqref{e30} $t_0=0$   as well as 
\be
\label{e32}
a=\alpha\epsilon^2, \quad  \alpha=const>0, 
\ee 
 and next  performing the limit $\alpha \rightarrow 0$ one obtains the instantaneous (abrupt) change of frequency   and the well-known form of  the corresponding  function  $\tilde b(t)$
\be
\label{e33}
\lim_{\epsilon\rightarrow 0} \tilde \omega_{II}(t)= \left\{
\begin{array}{cc}
\alpha & \quad \textrm{for} \quad  t\leq 0,\\ 
0& \quad \textrm{for} \quad   t>0;
\end{array}  
\right. \qquad 
\lim_{\epsilon\rightarrow 0} \tilde b(t)= \left\{
\begin{array}{cc}
1 & \quad \textrm{for} \quad  t\leq 0,\\ 
\sqrt{1+\alpha^2t^2}& \quad \textrm{for} \quad  t>0.
\end{array}  
\right.
\ee  
The non-zero ending value  can be obtained when we consider the following continuous profile 
\be
\label{e34}
\tilde{\tilde \omega}_{II}(t)= \left\{
\begin{array}{cc}
\omega_0\equiv \frac{a}{\epsilon^2+t_0^2} & \quad \textrm{for} \quad  t\leq t_0,\\ 
\omega_{II}(t)& \quad \textrm{for} \quad t_0< t\leq t_1,\\
\omega_1\equiv\frac{a}{\epsilon^2+t_1^2} & \quad \textrm{for} \quad  t_1< t.\
\end{array}
\right.
\ee  
Then
\be
\label{e35}
\tilde{\tilde b}^2(t)= \left\{
\begin{array}{c}
1  \quad \textrm{for} \quad  t\leq t_0,\\ 
b^2(t) \quad \textrm{for} \quad t_0\leq t\leq t_1,\\
\left(	b(t_1)\cos(\omega_1(t-t_1))+\frac{\dot b(t_1)}{\omega_1} \sin(\omega_1(t-t_1))\right)^2 +\frac{\omega_0^2}{\omega_1^2b^2(t_1)}\sin^2(\omega_1(t-t_1)) \quad \textrm{for} \quad  t_1<t;
\end{array}
\right.
\ee
where  the function $b(t) $ is given by \eqref{e27}.
For example,  for the  value $a=\sqrt 3 \epsilon$ and $t_0=-\epsilon=-t_1$  (i.e. a frequency  jump on the interval $[-\epsilon,\epsilon]$)  one   easily find  $b(t_1)=1$ and  $\dot b (t_1)=\frac{3}{4\epsilon}$).
Finally, let us   note that for the instantaneous change of the  frequency, from $\omega_0$ to $\omega_1$, at time zero    one has the know result
\be
\label{e36}
\tilde{\tilde b}(t)= \left\{
\begin{array}{cc}
1 & \quad \textrm{for} \quad  t\leq 0,\\ 
\sqrt{\cos^2(\omega_1t)+\frac{\omega_0^2}{\omega_1^2}\sin^2(\omega_1t)}& \quad \textrm{for} \quad  t> 0.
\end{array}  
\right.
\ee  
\par
Summarizing, in this section we have  analysed some special choices of the  time-dependent frequencies for which  the  corresponding classical and  quantum dynamics are   more  transparent since  the evolution is described by the elementary functions  (in contrast to the popular models  using, sometimes quite sophisticated,  special functions or singular frequencies).  This is especially relevant due to the fact that  the linear oscillator  appears as a building block  in many physical  problems  what, in turn,  involves its further processing;  we will  see it also   in   our investigations. 
\subsection{Dynamics of  entropic and information measures}
\label{s1c}
In this section, we analyse the  time evolution  of  entropic  and information measures in the case of  the quantum TDHO.  To this end let us  take the state  $\psi_n(x,t)$ which  at time  $t=t_0$ is an eigenstate of $\hat H(t_0)$, i.e. the initial conditions \eqref{e15} hold.    Then  the density function of the sate $\psi_n(x,t)$  reads 
\be
\label{e37}
\rho_n(x,t)=\frac{1}{2^n n! b(t)}\sqrt{\frac{\omega_0}{\pi}}H_n^2\left(\frac{x\sqrt{\omega_0}}{b(t)}\right)e^{-\frac{\omega_0 x^2}{b^2(t)}},
\ee
while  density of the  Fourier transform of $\psi_n(x,t)$ is given by the formula 
\be
\label{e38}
 \rho_n(p,t)= \frac{b(t)}{ 2^n n!\sqrt{\omega_0^2+b^2(t)\dot b^2(t)}}\sqrt{\frac{\omega_0}{\pi}}e^{-\frac{\omega_0b^2(t)p^2}{\omega_0^2+b^2(t)\dot b^2(t)}}H_n^2\left(\frac{\sqrt{\omega_0}b(t)p}{\sqrt{\omega_0^2+b^2(t)\dot b^2(t)}}\right).
\ee
 We start with  the position  and momentum Shannon entropies 
\be
\label{e39}
S_n^x(t)=-\int\rho_n(x,t)\ln\rho_n(x,t)dx, \qquad  S^p_n(t)=-\int\rho_n(p,t)\ln \rho_n(p,t)dp.
\ee
Substituting \eqref{e37} and  \eqref{e38} into \eqref{e39}  one  arrives at quite complicated terms related to   so-called Hermite entropies, see \cite{b11,b21};  moreover, both the  entropies  depend on $n$.  
  In contrast to this let us note that   we can directly compute  the increase of  the entropy for an arbitrary state $\psi_n$ and  it does not depend on $n$.
 In fact, by  direct calculations  we find that 
\be
\label{e40}
\begin{split}
\Delta S_n^x(t)\equiv S^x_n(t)-S_n^x(t_0)& = \ln(b(t)),\\
\Delta S_n^p (t)\equiv  S_n^p(t)- S_n^p(t_0)& = \frac 1 2 \ln\left(\frac{\omega_0^2+b^2(t)\dot b^2(t)}{\omega_0^2b^2(t)}\right);
\end{split}
\ee
thus the increase of  the entropy   depends on the function $b(t)$ only.    The same concerns    the joint entropy  $S^j_n(t)\equiv S_n^x(t)+ S_n^p(t)$  
\be 
\label{e41}
\Delta S^j_n(t)=\frac 12 \ln \left(1+\frac{b^2(t)\dot b^2(t)}{\omega_0^2}\right) .
\ee
Obviously, for the ordinary harmonic oscillator, i.e. $\omega(t)=const $,  we have $b(t)=1$ and all  the above  quantities vanish. 
\par 
Now, we show that a similar situation holds for the   R\'enyi entropies given by 
\be
\label{e42}
R^{\alpha,x}_n(t)=\frac{1}{1-\alpha}\ln\left (\int \rho^\alpha_n(x,t) dx\right),
 \ee 
 as well as  for their  momentum   counterparts $R^{\alpha,p}_n(t)$. 
 Indeed, by virtue of eqs.  \eqref{e37} and  \eqref{e38}, after straightforward computations,   we obtain  that in this case  the  increase does not   dependent on $n$ as well as  $\alpha$  
\be
\label{e42b}
\begin{split}
\Delta R_n^{\alpha,x} (t)\equiv R_n^{\alpha,x} (t)-R_n^{\alpha,x} (t_0)&=\ln(b(t)),  \\
\Delta   R_n^{\alpha,p}(t)\equiv  R_n^{\alpha,p} (t)- R_n^{\alpha,p}(t_0)&= \frac 1 2 \ln\left(\frac{\omega_0^2+b^2(t)\dot b^2(t)}{\omega_0^2b^2(t)}\right).
\end{split}
\ee
\par 
Let us now compute the  Fisher information for the states \eqref{e11} 
\be
\label{e43}
F_n^x(t)=\int (\partial_x\rho_n(x,t))^2\rho_n^{-1}(x,t) dx, \quad  F_n^p(t)=\int (\partial_p \rho_n(p,t))^2 \rho_n^{-1}(p,t) dp.
\ee
Substituting   \eqref{e37}  and \eqref{e38} into eqs. \eqref{e43} and   next using  the basic  properties of the Hermite polynomials    we get
\be
\label{e44}
F_n^x(t)=\frac{2\omega_0}{b^2(t)}(2n+1), \qquad  F_n^p(t)= \frac{2b^2(t)\omega_0}{\omega_0^2+b^2(t)\dot b^2(t)}(2n+1).
\ee
In consequence, we   find  that the product of the Fisher informations   is of the form 
\be
\label{e45}
F_n^x(t)  F_n^p(t)=\frac{4\omega_0^2(2n+1)^2}{\omega_0^2+b^2(t)\dot b^2(t)}.
\ee
These results fit into the considerations of Ref.  \cite{b5} where  it has  been shown that for the  real    wave functions the   product of the Fisher informations is greater than  or equal to  four; however, in the   general case there is no lower bound. In fact, taking $\omega_I$ and \eqref{e23}  we   see that  this product can be arbitrary small. However, in our case  we have a time-independent inequality  $F_n^x(t)  F_n^p(t)\leq  4(2n+1)^2$. 
Finally, let us  note that our results coincide with the    Stam  and Cram\'er-Rao  inequalities. In fact,   by virtue of \eqref{e16} and  \eqref{e17}, for  each $n\in\mN$   we have 
\be
\label{e46}
F_n^x(t)\leq 4\langle p^2\rangle_n (t),\quad  F_n^p(t)\leq 	4\langle x^2\rangle _n (t),
\ee
as well as
\be
\label{e47}
F_n^x(t)\geq\frac{1}{(\Delta_n^2 x)(t)}, \quad   F_n^p(t)\geq\frac{1}{(\Delta_n^2 p)(t)}.
\ee
\par 
Moreover, using eqs. \eqref{e40}  we can express the  $\Delta S_n^x(t)$ and $  \Delta S_n^p(t)$ in terms of the Fisher informations 
\be
\label{e48}
\Delta S_n^x(t)=\frac 12 \ln\left (\frac{F_n^x(t_0)}{F_n^x(t)}\right ), \quad \Delta   S_n^p(t)=\frac 12 \ln\left (\frac{F_n^p(t_0)}{ F_n^p(t)}\right ) .
\ee
In consequence, we observe that the Fisher-Shannon complexities 
\be
\label{e49}
C_n^{FS,x}(t)\equiv F^x_n(t)e^{2S_n^x(t)},\quad C_n^{FS,p}(t)\equiv F^p_n(t)e^{2S_n^p(t)},
\ee 
are constant in time  $C_n^{FS,x}(t)=C_n^{FS,x}(t_0)$ and $C_n^{FS,p}(t)=C_n^{FS,p}(t_0)$. 
In addition,    we have the following relation between the joint entropy and the Heisenberg uncertainty relation
\be
\label{e50}
\frac{e^{\Delta S_n^j(t)}}{2(2n+1)}\leq  (\Delta_n x)(t)(\Delta_n p)(t).
\ee
\par
\begin{figure}[!ht]
\begin{center}
\includegraphics[width=0.45\columnwidth]{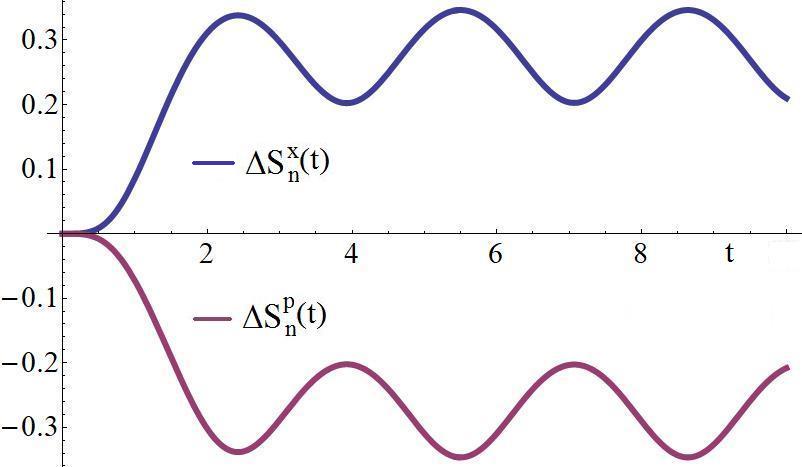} \hspace{1cm}
\includegraphics[width=0.45\columnwidth]{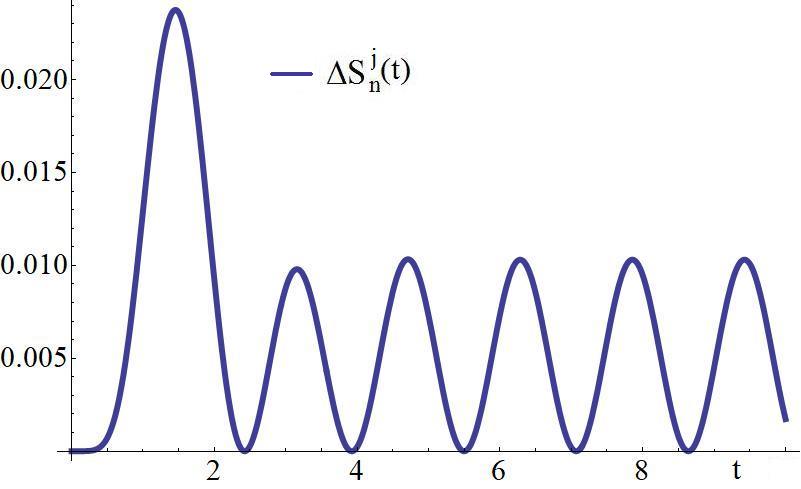}
\\
\end{center}
\caption{\small{The increase of   the position and momentum  Shannon  entropies as well as  the joint entropy    corresponding to the choice  \eqref{e20}  with  $a=\epsilon=1$.  }
\label{fig1}}
\end{figure}

\begin{figure}[!ht]
\begin{center}
\includegraphics[width=0.45\columnwidth]{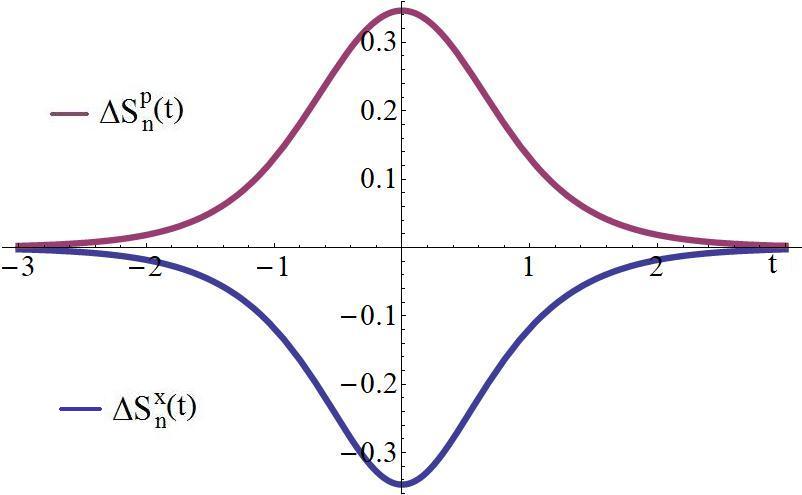} \hspace{1cm}
\includegraphics[width=0.45\columnwidth]{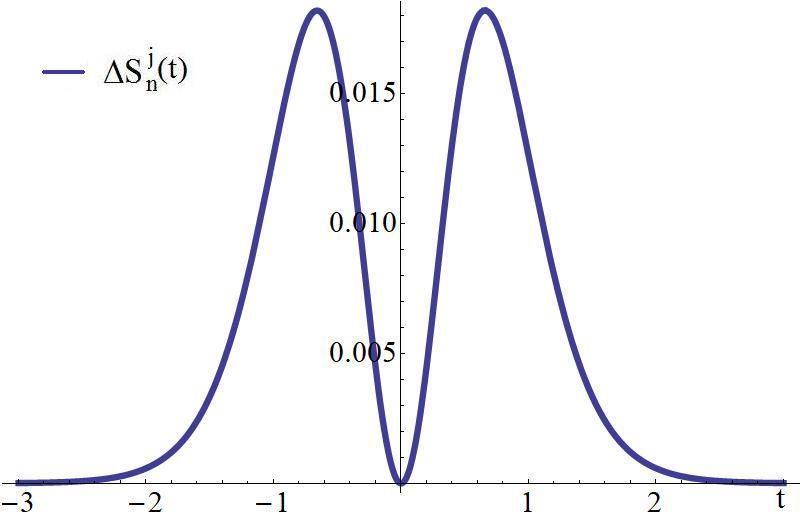}
\\
\end{center}
\caption{\small{The increase of the   position and momentum Shannon entropies as well as the  joint entropy   corresponding to choice  \eqref{e23}  with  $a=\epsilon=1$.  }
\label{fig2}}
\end{figure}

\begin{figure}[!ht]
\begin{center}
\includegraphics[width=0.45\columnwidth]{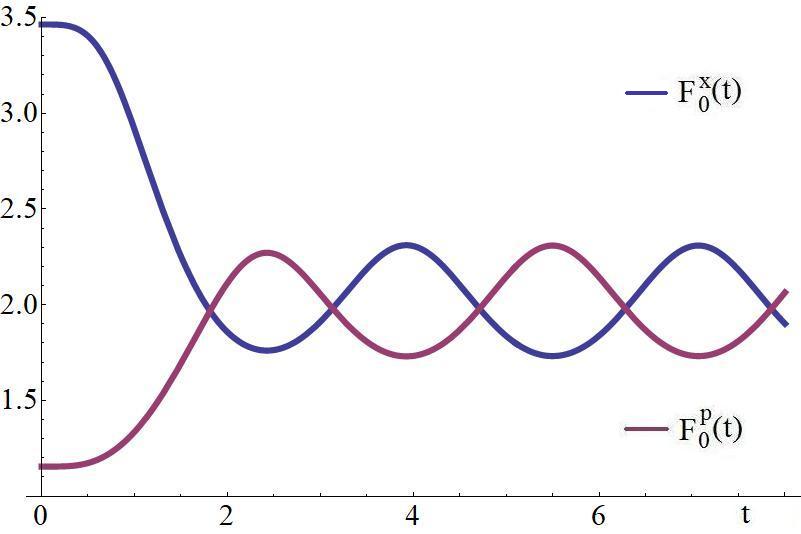} \hspace{1cm}
\includegraphics[width=0.45\columnwidth]{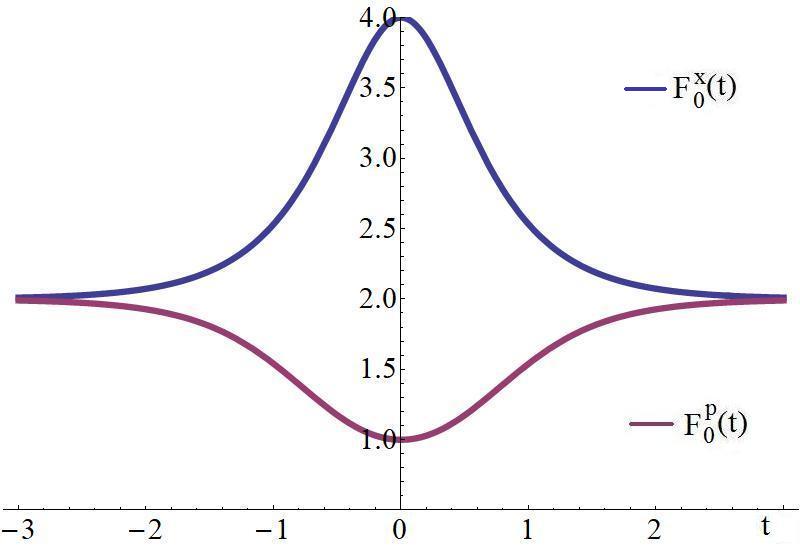}
\\
\end{center}
\caption{\small{The  dynamics of the Fisher information, for  $n=0$,   corresponding to \eqref{e20} and  \eqref{e23}, respectively,   with  $a=\epsilon=1$. }
\label{fig3}}
\end{figure}

 
At the end,  let us recall that for the frequencies presented is Sec. \ref{s1b} the  function $b(t)$ is   given  in terms of elementary functions.      In consequence, the increase  of  all discussed   entropies and Fisher informations  can be immediately obtained in these cases, for  illustration see Figs. \ref{fig1}-\ref{fig3}.
   \par Here, we only note that for the frequencies $\omega_{I}(t)$ with the initial conditions \eqref{e15}  at $t_0=-\infty$ (see eq.   \eqref{e23} for the function $b(t)$)   we obtain that $S^x_n(-\infty)=S_n^x(\infty)$  and analogously for other entropies;  in other words the  final entropies stabilize independently of  the history of the evolution (i.e.  the parameter $\epsilon$), see Figs. \ref{fig2} and \ref{fig3}.

\section{Uniform time-dependent magnetic fields}
\label{s2}
\subsection{The dynamics of entropies}
\label{s2a}
Let us consider the Hamiltonian of the charged  particle in the electromagnetic field defined by the potentials $\vec A$ and  $\Phi$ 
\be
\label{e59}
H=\frac{1}{2}(\vec p-\vec A)^2+\Phi,
\ee
(for simplicity we put $e=1$ and $m=1$). Then the Schr\"odinger equation  corresponding  to the above Hamiltonian (with the   potential in the Coulomb gauge) takes the form
\be
\label{e60}
i\partial_t\phi=-\frac 12  \triangle \phi+i(\vec A\cdot\vec \nabla)\phi+\frac 12 \vec A^2\phi +\Phi\phi.
\ee
The case of the  uniform and constant magnetic field  corresponds to the famous   Landau levels.  When we want  to relax that assumption and consider a   uniform but time-varying magnetic  field  $\vec B(\vec x ,t)=2\omega(t)\vec e_3$   the situation complicates,  since  due to the Faraday  law we have  also (in general) a time-dependent electric field $\vec E (\vec x,t)=-\dot \omega(t)|\bx|\vec e_\theta$.   However, in terms of potentials the situation can  be described in a similar way as in the constant case
\be
\label{e61}
\vec A(\vec x,t)=({\bf A}(\bx,t),0)=(-\omega(t)x^2,\omega(t)x^1,0), \quad \Phi=0,
\ee
where  the bold letters denote the two-dimensional vectors with the indices  $1,2$.  Substituting  the potential \eqref{e61}  into  \eqref{e60}  we get  that the third  coordinate  decouples and  the relevant dynamics is described by the equation     
\be
\label{e62}
i\partial_t\phi(\bx,t)=-\frac 12 {\bf \triangle}\phi(\bx,t)+\frac 12 \omega^2(t)\bx^2\phi(\bx,t)+\frac{p_3^2}{2}\phi(\bx,t)-i\omega(t)(x^2\partial_1-x^1\partial_2)\phi(\bx,t),
\ee
where $p_3$ is the conserved momentum related to the  decoupling constant.   The wave functions which form a basis of the solutions  of   eq. \eqref{e62}  have been obtained,   through  the polar coordinates, in Ref. \cite{b13}.  However, for such a choice of basis     it is difficult, in general,   to analyse their  entropic properties; this  is possible only for the  ground state  \cite{b13}. Here we apply a slightly  different  approach which enables us to  extend these considerations to   an  orthonormal  basis of states  as well   to construct the time-independent Fisher-Shannon complexities.  
 Namely,  by means of the  time-depended unitary transformation 
\be
\label{e63}
\psi(\bx,t)=e^{it\frac{p_3^2}{2}}\phi(R(t)\bx,t), \quad R(t)=
\left(
\begin{array}{cc}
\cos(\Omega(t))&\sin(\Omega(t))\\
-\sin(\Omega(t))&\cos(\Omega(t))
\end{array}
\right),
\ee
where $\dot \Omega(t)=\omega(t)$, we  reduce eq. \eqref{e62} to the following one
\be
\label{e64}
i\partial_t\psi(\bx,t)=-\frac 12 {\bf \triangle}\psi(\bx,t)+\frac 12 \omega^2(t)\bx^2\psi(\bx,t).
\ee
Thus, the wave functions 
\be
\label{e65}
\phi_{mn}(\bx,t)=e^{-it\frac{p_3^2}{2}}\psi_m(x^1\cos(\Omega(t))-x^2\sin(\Omega(t)))\psi_n(x^1\sin(\Omega(t))+x^2\cos(\Omega(t))),
\ee
where $\psi$'s are given by eq. \eqref{e11} form  an orthonormal basis for the solutions of the  transversal Schr\"odinger equation \eqref{e62}. 
\par Now,  imposing  on the function $b(t)$ the initial conditions \eqref{e15},  we  compute the change of the two-dimensional Shannon entropies  $S_{m,n}^{\bx}$ and $S_{m,n}^{\bp}$ of the states \eqref{e65}.
First,   using the fact that $\det(R(t))=1$, after some computations, we find  the increase of entropy  (from $t_0$ to $t$)  is of the form 
\be
\label{e66}
\triangle S_{m,n}^{\bx}(t)=2\ln(b(t)).
\ee 
 Moreover, performing the Fourier transform of the states \eqref{e65},  we  obtain  the momentum density $\rho_{m,n}(\bp,t)$   expressed as the product of the functions \eqref{e38} with  the arguments  $R^T(t){\bf  p}$. However, using  again the condition  $\det(R(t))=1$ we obtain that  
\be
\label{e67}
\Delta S_{m,n}^{\bp} (t) =  \ln\left(\frac{\omega_0^2+b^2(t)\dot b^2(t)}{\omega_0^2b^2(t)}\right) ,
\ee
 where $\omega_0=B(t_0)/2$.  Furthermore, we can  also find   the Fisher information for the discussed basis
 \be
 \label{e68}
F_{m,n}^{\bx}(t)=\frac{4\omega_0}{b^2(t)}(m+n+1), \qquad  F_{m,n}^{\bp} (t)= \frac{4b^2(t)\omega_0}{\omega_0^2+b^2(t)\dot b^2(t)}(m+n+1).
 \ee
Let us note that  for $m=n=0$  the above results  coincide with the ones obtained in Ref. \cite{b13}.  Finally,     the time-independent Fisher-Shannon complexities  can be also  constructed (cf. eqs. \eqref{e49})
\be
\label{e69}
F^{\bx}_{m,n}(t)e^{S_{m,n}^{\bx}(t)}=F^{\bx}_{m,n}(t_0)e^{S_{m,n}^{\bx}(t_0)},\quad F^{\bp}_{m,n}(t)e^{S_{m,n}^{\bp}(t)}=F^{\bp}_{m,n}(t_0)e^{S_{m,n}^{\bp}(t_0)}.
\ee 
\par
\begin{figure}[!ht]
\begin{center}
\includegraphics[width=0.45\columnwidth]{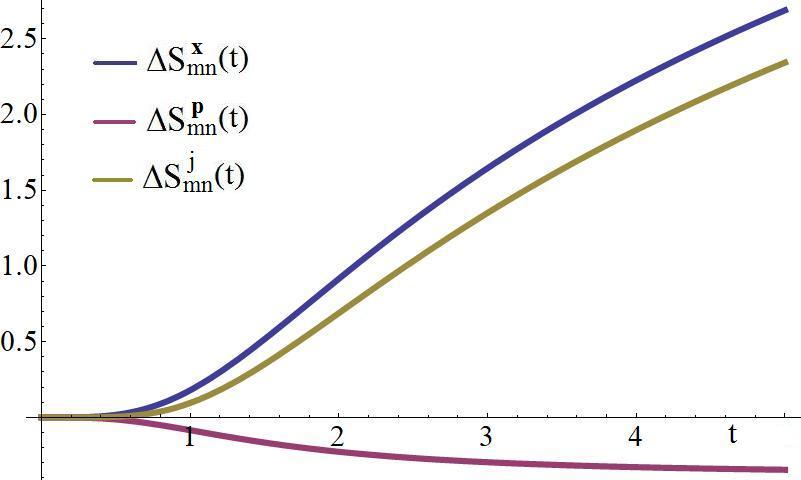} \hspace{1cm}
\includegraphics[width=0.45\columnwidth]{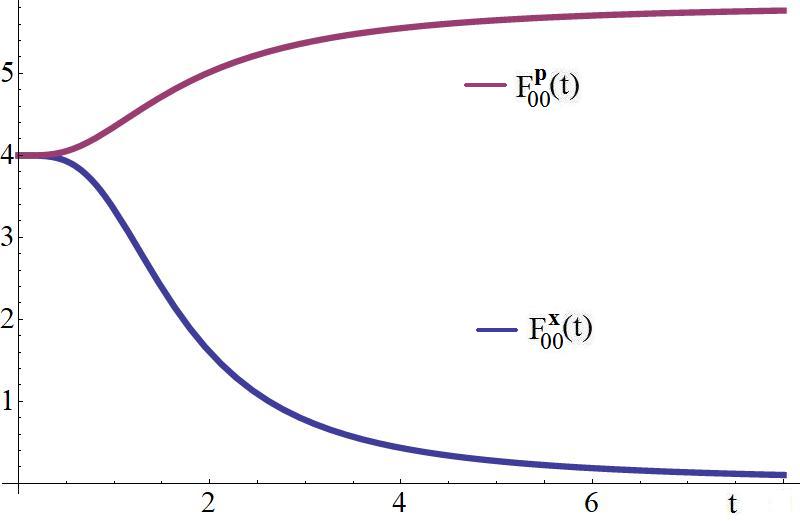}
\\
\end{center}
\caption{\small{The increase of  the  position, momentum  Shannon entropies and joint entropy   as well as the dynamics of the  Fisher information for the  electromagnetic field  \eqref{e61}  defined by \eqref{e25} with    $a=\epsilon=1$}
\label{fig4}}
\end{figure}

 
In view of the above  results we see that in order to analyse entropic relations  for  charged particle in the time-varying  electromagnetic field   \eqref{e61}, the explicit form  the function $b(t)$ is needed.   Such a situation  appears for the electromagnetic potential   defined by  $\omega(t) $ given by  $\omega_{II}$ (or $\omega_{I}$ with  $a=0$), see   \eqref{e25} (and \eqref{e18}). Then,  we have  electromagnetic pulses (with the uniform magnetic  field)  which disappear at plus/minus infinity.  Since for such a choice of   the electromagnetic  pulses the function $b(t)$ is an elementary one, see Sec. \ref{s1b},  thus  we immediately obtain  the  explicit time dependence of  the  discussed  properties  of the  entropy;   for illustration see  Fig. \ref{fig4}, where  we present the results for  the  electromagnetic pulse defined by the frequency \eqref{e25}.
\par
 It  is also  worth to notice that   putting  $a\sim \epsilon$   in $\omega_{II}(t)$ and next taking the limit   $\epsilon\rightarrow 0$   one obtains the Dirac delta  behavior    of the magnetic field $\vec B(t)\sim \delta(t)\vec e_3$ .
 Finally, we  will see in the next section that  the electromagnetic field defined by    $\omega_{II}(t)$   has a nice geometric interpretation.  
\subsection{Geometric analysis}
\label{s2b}
In the previous section we have seen that for  the  electromagnetic field  defined by  $ \omega_{II}(t)$  the description of the quantum dynamics of the  charged particle   simplifies. To better  understand this situation we come back to the classical level  where as we will  show below   such  a choice has a nice geometric interpretation. 
To this end let us recall that  the classical dynamics of the particle  governed by the electromagnetic potential $\vec A(\vec x,t)$   can be   embedded, by means of the so-called Eisenhart-Duval lift \cite{b13a}-\cite{b13e},   into the geodesic equation of the five-dimensional  spacetime
\be
\label{e70}
g=2dtdv+2dt\vec A(\vec x,t)\cdot d\vec x+d\vec x^2.
\ee
Namely, the geodesic equations corresponding to  the $\vec x$ coordinates reproduce the Lorentz equations with $\vec B$ and $ \vec E$ determined by $\vec A$; equation for the $v$ coordinate decouples while  $u$ is proportional to the affine parameter.  For the potential given by eq. \eqref{e61}   one arrives at the  four-dimensional metric 
\be
\label{e71}
g=2dtdv+2\omega(t)(x^1dx^2-x^2dx^1)dt+d\bx^2,
\ee
  for which only the one component  of the Ricci tensor is non-zero,  i.e. $R_{tt}(t)=(B^3)^2(t)/2=2\omega^2(t)$ (the metric has vanishing the scalar curvature and  describes a null-fluid solution to the Einstein equations).     Let us now  analyse the conformal Killing vectors of the metric \eqref{e71}.  Of course, one can write out the suitable equations describing the conformal fields. However, a more simpler way is based on the  following change of the coordinates  
  \be
  \label{e72}
  \bx=R(t)\tilde \bx,
  \ee
where $R(t)$ is given by    \eqref{e63}. Then, in the new coordinates, the metric takes the form 
\be
\label{e73}
g=2dtdv-\omega^2(t)\tilde \bx^2dt^2+d\tilde \bx^2,
\ee
i.e. it is a  conformally flat pp-waves \cite{b22}.  In consequence, for any function $\omega(t)$ the Lie algebra  of the conformal Killing vectors  of the metric $g$  is 15-dimensional;  however, the number of Killing, homothetic and proper conformal  fields depends on the choice of $\omega(t)$.  Now let us  recall  that for the null geodesics the conformal fields yield  constants of motion $J=Y_\mu\frac{dx^\mu} {d\tau}$ which, in turn,  in the generic case   (i.e. for the so-called chronoprojective fields)   can be projected onto constants of motion  of the  considered  classical dynamics (since the classical dynamics   is embedded into the geodesic equation);  for more details and further references  concerning this topic see   \cite{b13a}-\cite{b13e}.    Using the results of Ref.  \cite{b23} we verify  that  the  field  
\be
\label{e74}
Y=F(t)\partial_u-\frac{1}{4}\overset{..}{F}(t)\tilde{\bx}^2\partial_v+\frac{1}{2}\dot F(t)\tilde\bx\partial_{\tilde\bx},
\ee
where $F(t)$ satisfies equation  $\overset{...}{F}(t)+4\dot \omega(t) \omega(t) F(t)+4\omega ^2(t)\dot F(t)=0$,  is a  conformal vector field  of  the metric \eqref{e73} with   the conformal factor $f=\dot F/2$. Then, the three  independent solutions are:  $F(t)=b^2(t)$, $F(t)=b^2(t)\sin(c\tau(t))$ and $F(t)=b^2(t)\cos(c\tau(t))$ where $b(t)$ satisfies the EMP equation \eqref{e3} and $\tau(t)$ is given by  \eqref{e6}. 
Now, returning  to the initial variable  $\bx$  the conformal factor remains unchanged  but $Y$ takes the form 
\be
\label{e75}
Y=F(t)\partial_u-\frac{1}{4}\overset{..}{F}(t){\bx}^2\partial_v+\frac{1}{2}\dot F(t)\bx\partial_{\bx}-\omega(t)F(t)(x^1\partial_{2}-x^2\partial_{1}).
\ee
The field \eqref{e75}   implies the  integral of  motion  (for null geodesics) of the from
\be
\label{e76}
J=F(t)\dot v  -\frac{1}{4}\overset{..}{F}(t){\bx}^2-F(t){\bf A}^2(\bx,t)+\frac 12\dot F(t)\bx\cdot\dot\bx .
\ee
On the other hand, by virtue of the null condition of the geodesic one has $\dot v=-\frac 12 \dot \bx^2-{\bf A}\cdot \dot\bx$.  In consequence, the constant of motion $J$ can be  projected onto  the integral   of motion of the initial dynamics, i.e.  the Lorentz  equation for the electromagnetic field defined by \eqref{e61}; namely,  we  have
\be
\label{e77}
J=-\frac12 F(t)\dot \bx^2   -\frac{1}{4}\overset{..}{F}(t){\bx}^2-F(t){\bf A}^2(\bx,t)+\left(\frac 12\dot F(t)\bx-F(t){\bf A}(\bx,t)\right)\cdot\dot\bx;
\ee
now, $J$ depends   on $t$ and $\bx$ only.
\par 
Next, let us take the conformal field generated by $F(t)=b^2(t)$, where $b(t) $ satisfies eq. \eqref{e3}. Then the conformal factor  is of the form   $f(t)=b(t)\dot b(t)$ and the corresponding integral of motion takes the form
\be
\label{e78}
J=-\frac 12b^2(t)\dot\bx^2-\frac 12\left(\omega^2b^2(t)+\dot  b^2(t)+\frac{c^2}{b^2(t)}\right)\bx^2+(b(t)\dot b(t)\bx-b^2(t){\bf A}(\bx,t))\dot \bx .
\ee
In order to  make the meaning of $J$  more transparent let us note that it can be rewritten in the following form  
\be
\label{e79}
J=-\frac{1}{2}\left((b(t){\bf p}-\dot b(t)\bx)^2+\frac{c^2\bx^2}{b^2(t)}\right),
\ee 
where  ${\bf p}$ is the canonical momentum, i.e. ${\bf p}=\dot{\bf x}+{\bf A}$.  In summary,  the integral of motion  associated with such a conformal field corresponds to  the   Ermakov-Lewis invariant (cf. eq. \eqref{e13}); however, we should  keep in mind that it contains the canonical momenta (not the kinetic ones). 
\par 
Now, let us recall  \cite{b22} that among all proper conformal vectors  the most   interesting  seem the so-called special ones, i.e. when the conformal factor  $f$ satisfies   $\textrm {Hess}(f)=0$ (such a condition holds, for 
example, for  any conformal field of  the Minkowski spacetime or even   any vacuum solution to the Einstein equations). Following Ref.   \cite{b22} we have that the   metric given  by \eqref{e73}  admit   a special  conformal Killing vector if and only if  $\omega(t)=\omega_{II}(t)$. Moreover,   taking into account the  above considerations    this holds only for  the conformal Killing  field  \eqref{e74} defined  by
\be
\label{e80}
F(t)=b^2(t)=t^2+\epsilon^2.
\ee
In this case  the conformal factor is  $f(t)=t$  and $b(t)$  satisfies  the   EMP equation  \eqref{e3} with $c^2=a^2+\epsilon^2$.    In view of this and the considerations  presented in Sec. \ref{s1a}  (see the transformation \eqref{e7})   as well as eq. \eqref{e72}  we  immediately obtain  the explicit solvability of the  Lorentz equation  with  the potential \eqref{e61}  defined by $\omega(t)=\omega_{II}(t)$.  Finally, the integral of motion $J$,  defined by \eqref{e79}, in terms of ${\bf y}, \tau$  variables  corresponds to  the energy of  the two-dimensional, isotopic,  harmonic  oscillator with the frequency $c^2=a^2+\epsilon^2$.   
\section{Time dependent coupled oscillators}
\label{s3}
\subsection{Entanglement dynamics of coupled oscillators}
\label{s3a}
The aim of the present section  is to  show that the results of Sec. \ref{s1b}  can be  also useful in the  analysis of the entanglement entropy  for the system of two harmonic (in general with time-dependent frequencies)  oscillators coupled by  a time-dependent  parameter. More precisely, let us consider the Hamiltonian of  the form  
\be
\label{e81}
H(t)=\frac 12(p_1^2+p_2^2)+\frac 12\omega^2(t)\left((x^1)^2+(x^2)^2\right)+\frac 12k(t)\left(x^1-x^2\right)^2.
\ee
Then the transformation  ${\bf x}=R{\bf y}$ where $R$ is given  by  \eqref{e63} with $\Omega(t)=\pi/4$ transforms  the Hamiltonian \eqref{e81} into the following one
\be
\label{e82}
H_{\bf y}(t)=\frac 12(p_1^2+p_2^2)+\frac 12\left(\omega^2_1(t)(y^1)^2+\omega^2_2(t)(y^2)^2\right),
\ee
where  now  $p$'s denote  the canonical momenta associated  with $y$'s and $\omega_1^2(t)=\omega^2(t)+2k(t),\quad \omega_2^2(t)=\omega^2(t)$.    The frequencies $\omega_{1,2}(t)$ determine the  parameters of the initial Hamiltonian  \eqref{e81} as follows 
\be
\label{e83}
\omega(t)=\omega_2(t),\quad k(t)=\frac 12 (\omega_1^2(t)-\omega_2^2(t)).
\ee
\par 
The evolution  $\psi_0(\bx,t)$  of the ground state  $\psi_0(\bx,t_0)$ of the Hamiltonian  operator  $\hat H(t_0)$ as well as the  reduced density matrix $\rho_{0, red}(x^1,\tilde x^1,t)=\int \psi_0(x^1,x^2,t)\psi^*_0(\tilde x^1,x^2,t) dx_2$    can be  easily computed    when we  take into account the form of the  Hamiltonian  \eqref{e82} and next return to the $\bx$ variable. 
The final results is of the form (see \cite{b14})   
\be
\rho_{0,red}(x^1,\tilde x^1,t)=\frac{1}{\sqrt{\pi(\zeta(t)-\chi(t))}}e^{\chi(t)x^1\tilde x^1+ i((x^1)^2-(\tilde x^1)^2)\varphi(t)-\frac{\zeta(t)}{2}((x^1)^2+(\tilde x^1)^2)},
\ee
where 
\be
\label{e85a}
\varphi(t) =\frac{\dot b_1(t)}{4b_1(t)}+\frac{\dot b_2(t)}{4b_2(t)}-\frac{\frac{c_1}{b_1^2(t)}-\frac{c_2}{b_2^2(t)}}{\frac{c_1}{b_1^2(t)}+\frac{c_2}{b_2^2(t)}}\left(\frac{\dot b_1(t)}{4b_1(t)}-\frac{\dot b_2(t)}{4b_2(t)}\right),
\ee
 and $\zeta(t)>\chi(t)\geq 0$ are given by 
\be
\label{e85}
\zeta(t)=\frac{\left(\frac{c_1}{b_1^2(t)}+\frac{c_2}{b_2^2(t)}\right)^2+4\frac{c_1c_2}{b_1^2(t)b_2^2(t)}+\left(\frac{\dot b_1(t)}{b_1^2(t)}-\frac{\dot b_2(t)}{b_2^2(t)}\right)^2}{4\left(\frac{c_1}{b_1^2(t)}+\frac{c_2}{b_2^2(t)}\right)},
\quad 
\chi(t) =\frac{\left(\frac{c_1}{b_1^2(t)}-\frac{c_2}{b_2^2(t)}\right)^2+\left(\frac{\dot b_1(t)}{b_1^2(t)}-\frac{\dot b_2(t)}{b_2^2(t)}\right)^2}{4\left(\frac{c_1}{b_1^2(t)}+\frac{c_2}{b_2^2(t)}\right)},
\ee
while  the functions $b_{1,2}(t)$ satisfy the EMP equation \eqref{e3} with the frequencies $\omega_{1,2}(t)$ and the constants  $c_1=\sqrt{\omega^2(t_0)+2k(t_0)}$ and $c_2=\omega(t_0)$, respectively.   
  Then the R\'enyi  entropy $R^\alpha(t)=\ln(\tr(\rho_{0,red}^\alpha))/(1-\alpha)$  and  the  von Neumann  entropy  $S=\lim_{\alpha \rightarrow  1} R^\alpha$     of the  reduced density  $\rho_{0,red}$  can be computed,   see   Ref. \cite{b14} (see also  \cite{b24}) 
\be
\label{e84}
R^\alpha(t)=\frac{1}{1-\alpha}\ln\frac{(1-\xi (t))^\alpha }{1-\xi(t)}, \quad S(t)=-\ln(1-\xi(t))-\frac{\xi(t)}{1-\xi(t)}\ln\xi(t),
\ee
where $\xi(t)=\frac{\chi(t)}{\zeta  (t)+\sqrt{\zeta^2(t)-\chi^2(t)}}$.
In view of the above formulae we see that the entanglement entropy is directly determined by the solutions of the   two EMP equations with the  frequencies $\omega_{1,2}(t)$. 
\par
\begin{figure}[!ht]
\begin{center}
\includegraphics[width=0.45\columnwidth]{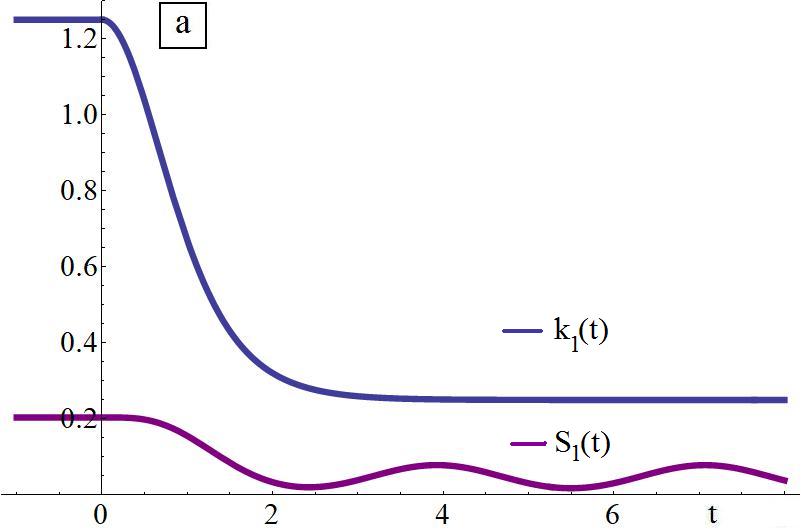} \hspace{1cm}
\includegraphics[width=0.45\columnwidth]{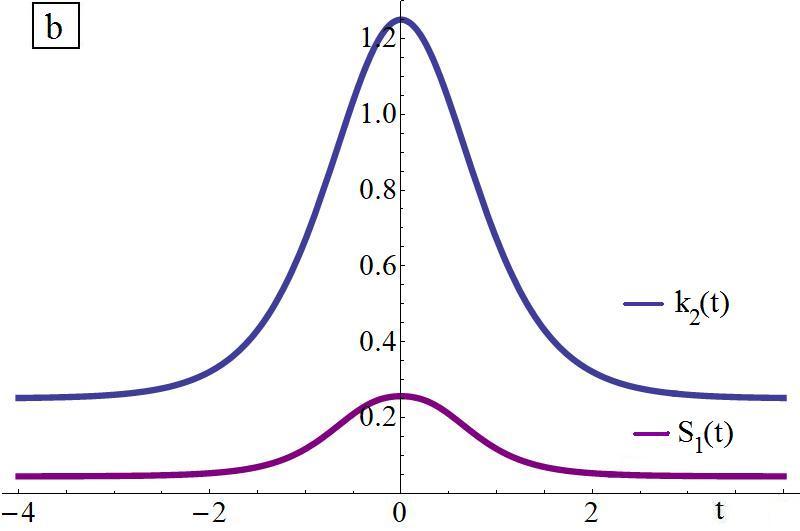}
\\
\end{center}
\caption{\small{The coupling parameter and the  corresponding entanglement entropy  a) $k_1(t)$ b) $k_2(t)$,  $a=\epsilon=1, \ \omega_2^2=1/2$.   }
\label{fig5}}
\end{figure}

 
Eqs. \eqref{e85} and \eqref{e84}   enable  us to analyse  directly the  dynamics of  various  entropies provided we have explicit solutions to the EMP equation.  Such a situation holds for    abrupt profiles and it was presented  in   Refs. \cite{b14,b25b};  here, we  complete these results  by considering  examples with the continuously changing parameters; in particular, the ones for which the entropy stabilizes.    
 To this end, we  use the exact examples  presented in Sec. \ref{s1b} to describe dynamics of the entropies for different  forms of the coupling $k(t)$ and frequency $\omega(t) $ appearing in the  Hamiltonian \eqref{e81}.  
Namely, taking $\omega_2^2(t)=\omega_2^2=const<a^2$ (thus  $b_2(t)=1$)   and $\omega_1(t)=\tilde\omega_{I}(t)$ (with $\tilde b_1(t)$  given by  eq. \eqref{e31}), by virtue of \eqref{e83}, we obtain the two  harmonic oscillators both with the frequency  $\omega_2$   coupled by  the decreasing (quenched)  function  $k_1(t)$, for  which the values  start at $(2/\epsilon^2+a^2-\omega_2^2)/2$ and end at $(a^2-\omega_2^2)/2$ (see Fig. \ref{fig5}a).  A similar situation holds when we take    $\omega_2(t)=\omega_2=const<a/\epsilon^2$ and   $\omega_1(t)=\tilde{\tilde{\omega }}_{II}(t)$ with $t_0=0$.
Another example is given by  $\omega_1 (t)=\omega_{I} (t) $  and   the initial condition at $t_0=-\infty$; then we obtain again the  two harmonic oscillators  coupled by the  parameter $k_2(t)$;  however this time the coupling  starts and ends at $(a^2-\omega_2^2)/2$ at minus/plus infinity  and  attains the maximal value at $t=0$ (see Fig. \ref{fig5}b). Let us stress that in this case  for the  function $b_1(t)$ we have $b_1(-\infty) =b_1(\infty) $  (see eq. \eqref{e23})  thus  the final  value of the entanglement entropy  (see eqs. \eqref{e84} and \eqref{e85}) stabilizes (there is no oscillatory behavior at $t=\infty$)   and it is  equal to the  initial  one (see also Fig. \ref{fig5}b).  Such a  situation  is quite different from the  generic oscillatory behavior, cf.  \cite{b14,b25b};   moreover,  it  holds  independently of the history of evolution  (i.e. the parameter $\epsilon$). 
\par
Of course,  one can take other combinations of frequencies $\omega$'s presented in Sec. \ref{s1b}  to obtain various forms of  the time-dependent harmonic oscillators  coupled by time-dependent parameters; for example  taking   $\omega_1(t)=\omega_{II}(t)$ and $\omega_2(t)=\omega_{I}(t)$  both with the parameter $\epsilon=1$ (implying $k\geq 0$)  or changing the the values $a,\epsilon$ of  the frequencies.   More generally,  those  frequencies   can be  directly applied to the   Wigner distribution functions,   entropies or  uncertainty relations of many integrating time-dependent harmonic oscillators   where more functions $b(t)$'s are involved (also for excited states), see \cite{b25a}-\cite{b25e}  for more details. 
\subsection{Quantum decoherence}
\label{s3b}
\par
One of the outstanding features of the quantum mechanics   is   the superposition principle.  However, this feature is very  fragile and its loss can  lead to  many  problems, in particular the ones related to     quantum computations   and computers (e.g. construction  of  quantum memory).   Usually the  loss of  quantum coherence, or, in general, the occurrence of the  quantum-classical transition,  is  related to the  dissipative  interaction of the system with  the environment. In consequence, such problems  are quite complicated and involve   the framework  of the quantum open systems (such as master equation) \cite{br1a,br1b}.  Moreover,  various,  inequivalent, measurements of decoherence (classicality)  have been proposed;  the main  ones are  based on the off-diagonal elements of the  density matrix.     Let us have a look on these issues in the context of  the results obtained in previous sections.
\par 
First, let  us  recall that for the density matrix $\rho(x,\tilde x,t)$  in the Gaussian form   a measure of the degree of  decoherence (related to the damping of the off-diagonal elements of  
$\rho$) is given by the following formula  
\be
\label{er1}
\delta_{QD}(t)=\sqrt{\frac{\Gamma_c(t)}{\Gamma_d(t)}},
\ee  
where  $\Gamma$'s are  coefficients appearing in  the density matrix when it is expressed in terms of  the variables $x_c=(x+\tilde x)/2$ and $x_d=(x-\tilde x)/2$, i.e. 
\be
\label{er0}
\rho(x_c,x_d,t)=\exp(-\Gamma_c(t) x_c^2-\Gamma_d(t)x_d^2-\Gamma_e(t)x_cx_d+\textrm{linear terms}).
\ee
On the other hand, there is a second, independent, measure  of the classicality of the quantum system; namely,    the classical correlation  $\delta_{CC}$  which measures the sharpness ($\delta_{CC}\ll 1$)  of the  Wigner function around the classical trajectory;  for the density matrix  \eqref{er0} it  is given  by  
\be
\delta_{CC}(t)=\frac{\sqrt{\Gamma_c(t)\Gamma_d(t)}}{|\Gamma_e(t)|}.
\ee
For more details concerning $\delta_{QD}$ and  $\delta_{CC}$  we refer  to  Refs. \cite{br1a,br1c,br1d}; here we only  note that $\delta_{QD}$  can be  expressed by  the purity of  $\rho$, i.e.  $\delta_{QD}= \tr(\rho^2)$ (thus $\delta_{QD}\leq 1$ and it is  representation invariant), while $\delta_{CC}$ is a  dimensionless quantity. 
\par 
To begin with, let us apply the above  measures to the TDHO \eqref{e1}.  For  the state $\psi_n$, see   \eqref{e11}, with $n=0$ we  obtain 
\be
\label{er6}
\delta_{QD}^0(t)=1, \quad \delta_{CC}^0(t)=\frac{c}{2b(t)\dot b(t)}.
\ee
Thus  there is no quantum decoherence,   though the entropies  change in time  (see eqs.  \eqref{e40} and \eqref{e41}). In this case   the change of the entropy is related to the second measure of the classicality. In fact, by virtue of eq. \eqref{e41}   (for $n=0$ and $c=\omega_0$) we obtain the following  relation   between the joint entropy and classical correlation 
\be
\label{er7}
\Delta S_0^j(t)=\frac 12 \ln\left(1+\frac{1}{4\delta_{CC}^2(t)}\right);
\ee
it can be also easily inverted.  
 For  the  ordinary harmonic oscillator and ground state ($b(t)=1$)  we obtain, as expected,  infinite $\delta_{CC}^0$ and the constant entropy.  For  the time-dependent case  $\omega(t)$  with  the final frequency equal to zero  (see eq. \eqref{e19}  and \eqref{e33} for  abrupt quench) the classical correlation can be arbitrary small for sufficiently large time,    while $\delta_{QD}^0$ remain unchanged.  
\par 
According to the general  belief  the decoherence phenomena appear when a  quantum system interacts  with some environment \cite{br1a,br1b}. In general, such an approach involves a more complicated analysis. However, in   the case of  two oscillators  coupled to each other  the first can be treated  as the system under consideration and the second one as the environment; then  the reduced density matrix can be used to  describe    the influence  of the  surroundings on the system. In view of this let us  consider the Hamiltonian  given by  \eqref{e81}. By means of eq. \eqref{er1},   or alternatively  computing the trace of $\rho^2_{0,red}$,   we find that 
\be
\label{er3}
\delta_{QD}^0(t)=\sqrt{\frac{\zeta(t)-\chi(t)}{\zeta(t)+\chi(t)}}=\frac{1-\xi(t)}{1+\xi(t)}.
\ee
In view of this the quantum decoherence emerges  and  it is time-dependent (in contrast to the single, noninteracting,  TDHO studied above). In terms of the solutions $b_1$ and $b_2$ of the EMP equation  \eqref{er3} takes the form 
\be
\label{er2}
\delta_{QD}^0(t)=\frac{2{\sqrt {c_1c_2}b_1(t)b_2(t)} }{\sqrt{(c_1b_2^2(t)+c_2b_1^2(t))^2+(\dot b_1(t)b_2(t)-\dot b_2(t)b_1(t))^2b_1^2(t)b_2^2(t)}}.
\ee  
The formula \eqref{er2} simplifies for  two   harmonic oscillators   coupled by time dependent parameter $k(t)$  (since  $b_2(t)=1$);  then we can simply use the examples  discussed  in  Sec. \ref{s3a} to  analytically analyse the degree of  quantum decoherence.  In the simplest case of   two   harmonic oscillators   coupled by a  constant parameter,   $\delta_{QD}^0$  reduces to a  constant which coincides with  observation made in Ref. \cite{br4a}. Finally, it is also worth to notice that the   relation \eqref{er3}  can be  inverted $\xi(t)=(1-\delta_{QD}^0(t))/(1+\delta_{QD}^0(t))$ and   then    various entropies (see eqs. \eqref{e84}) can be expressed in terms of  quantum decoherence.     
\par 
For the second measure, i.e.  the classical correlation,  we get
\be
\delta_{CC}^0(t)=\frac{\sqrt{\zeta^2(t)-\chi^2(t)}}{4|\varphi(t)|},
\ee
where $\zeta(t),\chi(t)$ and $\varphi(t)$  are given by eqs. \eqref{e85a} and \eqref{e85}. For the  the harmonic oscillators and constant  coupling  $\delta_{CC}^0(t)$  is infinite;  this is in contrast with the  time-dependent  case where  it is usually  finite, enhancing  in this way the classicality of the system (this can be  again easily seen using the results presented in Sec. \ref{s3a}).  These considerations   suggest  that  the  time-dependent   coupling (interaction)  of  the system with  environment  can lead to more serious destruction of quantum properties. 
\par
To conclude our  investigations  let us note that the above results  can be also   useful in other models and thus in further studies of  the  decoherence  phenomena. 
To this end let us consider  the  TDHO  driven by an external time-dependent force, i.e. 
\be
\label{er4}
\overset{..}{x}(t)=-\omega^2(t)x(t)+f(t).
\ee
It turns  out that   the quantum counterpart of \eqref{er4} can be reduced to  the force-free case \cite{br5}.
Namely, the  solution of the Schr\"odinger  equation  corresponding to  \eqref{er4} is of the form 
\be
\label{er5}
\phi(x,t) =\psi(x-e(t),t)e^{i\dot e (t)(x-e(t))+i\int Ldt},
\ee
where $\psi$ is a solution of the force-free Schr\"odinger equation, $L$ is the classical Lagrangian for eq. \eqref{er4}   and   $e(t)$ is a  solution to the classical equation of motion \eqref{er4}. 
 In view of this  the  frequencies $\omega(t)$  for which explicit solutions of the TDHO  are known (see Sec. \ref{s1})  can be  also very   useful   when  $f$ is  not  equal to zero.  Such a situation appears, for example,  in  the study of  the  entropies and   measures of  the classicality mentioned above.
 \par  In fact,    by means  of   eq. \eqref{er5} we find, after straightforward  calculations, that  all entropies $S^x$, $S^p$  and $S^j$ reduce to the ones for the force-free case (in particular  they are described   by the function $b(t)$ for the states  $\phi_n$  corresponding to   $\psi_n$ defined  in Sec. \ref{s1}). Moreover, the  $\delta_{QD}^0$ and $\delta_{CC}^0$  for $\phi_0$   are given also by \eqref{er6}  for the TDHO with an arbitrary driven force; in consequence, the relation \eqref{er7} remains   valid and there is no  decoherence. However,  the discussed  driving model \eqref{er4}  has an  interesting modification    when the external force change randomly \cite{br6}. Then even in the case of the  harmonic oscillator the joint  entropy changes in time \cite{b12d}; this  suggests that  for a random driving  force    $\delta_{QD}^0$  can also change with time;  however,  this involves more careful  analysis of the  ensemble average of  the density matrix. 
\section{A quantum quench for  non-relativistic fermions}
\label{s4}
One of the examples of  basic  phenomena  where the TDHO can be useful are the ones related to non-equilibrium processes  for which    time-dependent parameters appear.  Such  systems are modeled by  quantum fields   subjected  to a quench.  It turns out that in  the relativistic   theory some universal  phenomena emerge for the (a)diabatic regime; for example,the  Kibble-Zurek scaling, for review see  \cite{b26}.  On the other hand, from the experimental point of view the non-relativistic theories are also interesting.  In consequence, the question  arises whether a similar behavior emerges  for non-relativistic systems.   Such a problem has been recently discussed  in Ref.  \cite{b17} for  the  system of many   mutually non-interacting non-relativistic fermions in a  harmonic trap  (important in the context of  cold atom physics).   In particular,  it has been shown   that the description of the dynamics  of basic quantities  for  such a system is related to  the 	EMP equation \eqref{e3}; for example,  for  the expectation value $\langle \hat O\rangle(t)$  of the  operator 
\be
\label{e51}
\hat O(t)= \int x^2\hat \Psi(x,t)^\dag \hat\Psi(x,t)dx,
\ee
between  the ground   "in" states    (defined at early times) we have 
\be
\label{e51b}
\langle \hat O\rangle(t)\sim b^2(t).
\ee
 Furthermore,   the entanglement entropy  in a given  finite subregion    for  large number of fermions can be also  related to $b(t)$, it is proportional to the area  of the region and $b^{-1}(t)$   (for more details we refer to   \cite{b17}, in particular see formulae 3.8 and 7.8 therein). 
\par 
     A typical situation is when  the initial Hamiltonian is gapped, while the frequency  crosses or approaches a critical point where the gap vanishes.  
The last possibility, where the initial frequency  decreases to zero at late times (release from the harmonic trap) is  called the ending protocol. 
Such a protocol  has been  analysed  in  Ref. \cite{b17} by means of $\omega^2(t)\sim 1-\tanh(t)$.
However,  for such a choice the general  solution of the EMP equation \eqref{e3}  is a quite complicated special function;  in consequence, the analysis of  (a)diabatic regions quite   involved.
 In contrast,  for the quenched protocol modeled   by  the frequency \eqref{e22} with $a=0$   or  \eqref{e30} with $t_0=0$,  the function $b(t)$ is an  elementary one and thus the   (a)diabtic analysis  of the  expectation value \eqref{e51b}  as well as the estimation of the entanglement entropy of a subregion become more accessible.  Let us see this in more detail.  
\par 
 In order to  control both the  slow and fast quenches  in the  neighborhood of the critical point  and to  include the instantaneous quench, see eq. \eqref{e33}, we will use  $\tilde\omega_{II}(t)$ with  $a=\alpha\epsilon^2$ and  $t_0=0$.    Then  the function $b(t)$ for $t\geq 0$   can be expressed as follows
\be
\label{e52}
b^2(s)=\frac{s^2+1}{2(1+\beta^2)}(\cos(2\sqrt{1+\beta^2}\tan^{-1}(s))+1+2\beta^2),
\ee 
where $s=t/\epsilon$ and  $\beta=\alpha\epsilon$.  Since the frequency $\tilde\omega_{II}(t)$  approaches   the critical  point at plus infinity,  the adiabaticity  breaks down at some time $t_c$ (the so-called Kibble-Zurek time).   In our case  the Landau criterion 
\be
\label{e53}
\frac{|\dot\omega(t)|}{\omega^2(t)}\sim 1,
\ee
implies $t_c\sim \alpha\epsilon^2/2$, thus we   put     $s_c=\beta$.  On the other hand,   the adiabatic approximation $b_{ad}^2(t)=\omega_0/\omega(t)$  yields
\be
\label{e54}
b_{ad}^2(s)=1+s^2.
\ee 
In consequence, by virtue of \eqref{e51b}  we obtain the  approximative relation $\langle \hat O \rangle(s_c) \sim b_{ad}^2(s_c)= 1+\beta^2\sim \beta^2$ (in the  slow regime $\beta\gg 1$).   This result can be easily refinement by  substituting $s=s_c=\beta$ into eq. \eqref{e52}
\be
\label{e55}
b^2(s_c)=\beta^2+\cos^2(\sqrt{1+\beta^2}\tan^{-1}(\beta)).
\ee
Thus  there is an  additional term which only for $\beta\gg 1$ (slow regime)  can be skipped and then   the scaling is consistent with the Kibble-Zurek argument. 
\par
\begin{figure}[!ht]
\begin{center}
\includegraphics[width=0.45\columnwidth]{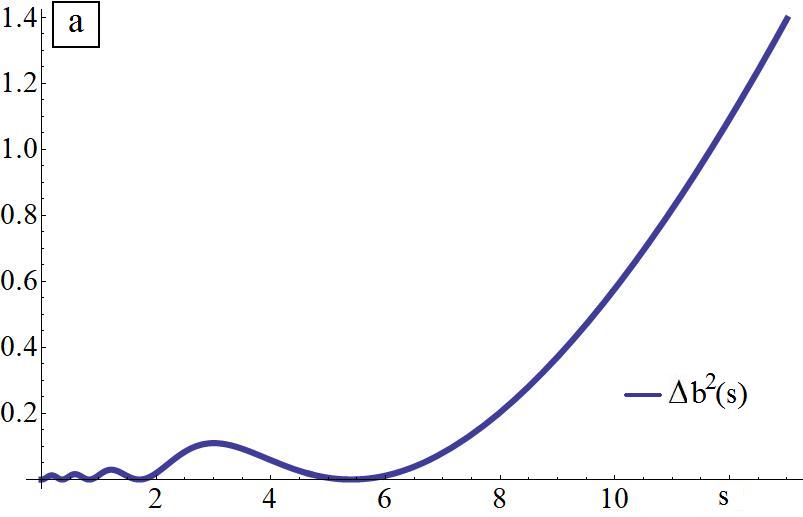} \hspace{1cm}
\includegraphics[width=0.45\columnwidth]{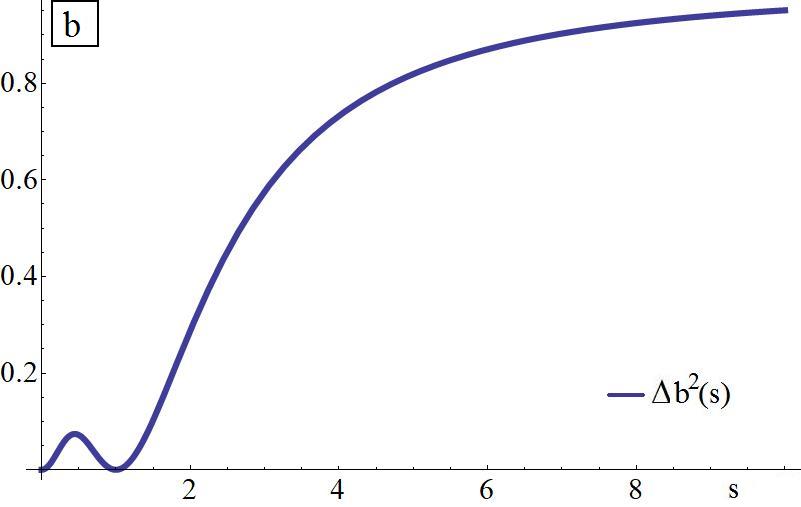}
\\
\end{center}
\caption{\small{ The function $\Delta b^2(s) $ a)  for   $\beta=9$,  b) for  $\beta=\sqrt{15}$ }
\label{fig7}}
\end{figure}

 
In general,  the  difference between the adiabatic solution $b_{ad}^2(s)$ and the exact  one is of the form  
\be
\label{e56}
\Delta b^2(s)\equiv b_{ad}^2(s)-b^2(s)=\frac{s^2+1}{1+\beta^2}\sin^2(\sqrt{1+\beta^2}\tan^{-1}(s)).
\ee
From eq. \eqref{e56}  follows that for a sufficiently large value of $\beta$ (i.e. $\beta^2>3$)   the function $\Delta b^2(s)$  vanishes  at some initial points  (for larger $\beta$ we have more points)  and  thus  $b(s)$  is close to the    adiabatic solution;  for   large value of $s$ ($s\gg s_c$) the function   $\Delta b^2(s)$  increases to infinity (see  also  Fig. \ref{fig7}a). However,  for special values, i.e.  $\beta=\sqrt{4k^2-1}$ where $k\in\mN$,    we have 
$\lim_{s \rightarrow \infty}\Delta b^2(s)=1$. In particular, for  $k=1$ ($\beta=\sqrt 3 $)  we have that 
\be
\Delta b^2(s)=\frac{s^2}{1+s^2},
\ee
thus  $0\leq \Delta b^2(s)<1$; despite   a quite small value of  the parameter   $\beta$ the  distance   from the adiabatic solution remains bounded  for all times (see also Fig. \ref{fig7}b for  $k=2$, i.e. $\beta=\sqrt{15}$). 
\par 
Now, let us have  a look on the fast regime and late times. More precisely, we assume that $\beta\ll 1$ and $t\gg \beta$  (equivalently  $s\gg1$). Then expanding \eqref{e52}  with respect to $s$ we have
\be
\label{e57}
b^2(s)=\frac{1}{2\lambda^2}\left(s^2(2\lambda^2+\cos(\pi\lambda)-1)+2s\lambda\sin(\lambda \pi) +(2\lambda^2-1)(1-\cos(\lambda \pi))+O(\frac{1}{s})\right).
\ee
Now,  taking into account that  $\lambda^2=1+\beta^2$ and expanding eq.   \eqref{e57} up to $\beta^2$  we have 
\be
\label{e58}
b^2(s)\simeq s^2\beta^2-s\frac{\pi\beta^2}{2} +\beta^2 +1+O(1/s)\equiv \alpha^2t^2+1-t\frac{\pi\alpha^2\epsilon }{2} +\alpha^2\epsilon^2 + O(1/t).
\ee
Comparing it with  eq.   \eqref{e33}, we see that the two last terms of  \eqref{e58}   describe  corrections to the  abrupt quench for which  the frequency suddenly changes from $\alpha$ to zero. 
\par
In summary, using $\tilde \omega_{II}(t)$ we  can simplify  the analysis  of quenched processes  which exhibit critical points. In consequence, some basic quantities,   such as expectation values and   the entanglement entropy of a  finite subregion for  large number of fermions  takes a more accessible form. 
\section{Summary and outlook}
\label{s5}
In this work we  have analysed dynamical aspects of  information-theoretic  and entropic   properties  of  various time-dependent  quantum systems.  We started with  the harmonic oscillator with the  time-dependent frequency.      In this case we showed that the   increase of the  position and  momentum Shannon (R\'enyi) or joint   entropies  of the states which initially are  eigenstates  of the  instantaneous  Hamiltonian depend only on the solution of the EMP equation; the same concerns the dynamics of the Fisher information of those states.   These results allowed us to examine the  Cram\'er-Rao inequalities  as well to find the explicit relation between the Fisher information and the increase of the Shannon entropy.  As a consequence,  we found that the Fisher-Shanon complexities are time independent quantities for such a basis.    Next, we have  shown  that a similar situation holds for  a suitable choice of the basic wave functions of  the charge particle in the uniform and time-dependent magnetic field (supplemented by a electric field). 
\par
In order to  illustrate the results we considered  some examples  of frequencies for which  the solutions of the  EMP equation  are elementary ones; in consequence, all the  mentioned dynamical relations take immediately the explicit forms.  Moreover, by means of the Eisenhart-Duval lift, we showed that some conformal Killing vectors imply the integrals of the  (geodesics) motion  which, in turn,   naturally lead to  the Ermakov-Lewis invariants  for  the considered electromagnetic fields. In particular, we  have shown that the existence of the special conformal vector  implies solvability  for the   one of those fields.
\par
Next, we  have explicitly worked out   the entanglement entropy of the  harmonic (in general, with time-dependent frequencies) oscillators coupled by a  continuous  time-dependent parameter.  In particular, we showed that for  a special form of the coupling parameter the final  value of the  entanglement entropy stabilizes  (independently of the history of evolution).  
We  also showed  that the above results and analytical examples  can be  useful for the study  of the quantum-classical  transition. To this end  we have examined  two independent  measures of the classicality and their relation with the entropy;   in particular,  we considered  some aspects of  the quantum  decoherence  which plays  the relevant role in  quantum information processing  (technology).  
\par 
In the last part of the work we have  studied in some detail the   behavior of quantum quenches (in the presence of the critical points) for the case of   mutually non-interacting non-relativistic fermions in a harmonic trap. In particular,  we  explicitly analysed  the scaling behavior  of the basic expectation values in the context of  the   Kibble-Zurek argument and adiabatic limit.  Moreover,  the discussed  exact solutions of the EMP equation  yield    direct  description of  the entanglement entropy  of a given subregion for large number of fermions.
\par
The results obtained  can  serve as a starting point for further considerations. Let us point out a few of them.  First,   following Sec.  \ref{s1} we can consider  further  information-theoretic   aspects of  quantum systems  such as the Tsallis entropy or the LMC shape complexity \cite{b27b}  and/or introduce time-dependent mass.  On the other hand, in view of  Refs.  \cite{b14,b25c}  the examples from Sec. \ref{s1b}  can be  directly used to illustrate, in the exact analytical form,   the time-dependent von Neumann and R\'enyi entropies for  a system of  many coupled oscillators, following a continuous quench. Moreover,  they can be applied   to the Wigner distribution functions and/or  anisotropic oscillators  \cite{b25b,b25cc,b25d}.  They can be  also useful  in a more general  frameworks of the  perturbative theory \cite{b28c} (note that we can compute the explicit form of propagator for the  discussed frequencies).   Moreover,   it would be interesting to compare   the results with the numerical methods based on the so-called Gaussian state approximation (such an approach has been quite recently applied for  instantaneous quenches \cite{b28d}) as  well as 
  to analyse the  time-dependent case of vanishing frequencies  (UV  divergences) in the spirit of the work \cite{b28e}. 
Finally,  note that the considerations from the last  section fit perfectly into the recent studies  \cite{b29} of quantum quenches of   the $c = 1$  Matrix Model  which, in turn, is  related to   the description of two-dimensional string theory with the  time-dependent string coupling.

  \end{document}